# Period-Doubling Route to Chaos and Intermittency in a Hybrid Rössler Model

Mehmet Onur Fen[1,*] and Fatma Tokmak Fen[2]

[1] Department of Mathematics, TED University, 06420, Ankara, Turkey

[2] Department of Mathematics, Gazi University, 06560, Ankara, Turkey

[*]Corresponding Author E-mail: onur.fen@tedu.edu.tr

**Abstract**

A Rössler model perturbed with a piecewise constant function is investigated. The perturbation function used in the model is constructed by means of the logistic map. In the absence of the perturbation the system is assumed to possess two equilibrium points one of which is linearly stable. The occurrences of period-doubling cascade and intermittency are numerically investigated. Extensions of the aforementioned phenomena among coupled Rössler systems are also shown. Our results reveal that discontinuous perturbations are capable of generating continuous chaos.

**1. Introduction and Preliminaries**

Researches on chaotic dynamics have gained popularity among scientists starting with the papers of Lorenz [1, 2]. Nowadays, chaos has applications in various fields such as mechanics, electronics, geology, biology, neural networks, image encryption, and wireless propagation [3]-[9]. For that reason it is an indispensible part of studies related to dynamical systems.

In 1976, Rössler [10] proposed the following system consisting of three differential equations:

$$x' = -y - z$$
$$y' = x + 0.2y \quad \quad (1)$$
$$z' = 0.2 + z(x - 5.7).$$

It was demonstrated in [10] that system (1) possesses chaotic oscillations, and its flow gives rise to the formation of a horseshoe map. One of the differences of Rössler model (1) compared to Lorenz system [2] is that there is only one nonlinear term in the former whereas two such terms take place in the latter.

In general, the Rössler system can be described by the system

$$x' = -y - z$$
$$y' = x + ay \quad \quad (2)$$
$$z' = b + z(x - c),$$

where $a, b, c$ are positive real numbers [10]-[12]. According to the results of [11], for $c^2 > 4ab$, system (2) admits the equilibrium points

$$P_1 = (-aq_1, q_1, -q_1) \tag{3}$$

and

$$P_2 = (-aq_2, q_2, -q_2), \tag{4}$$

where $q_1 = \frac{1}{2}\left(-\frac{c}{a} - \frac{\sqrt{c^2-4ab}}{a}\right)$ and $q_2 = \frac{1}{2}\left(-\frac{c}{a} + \frac{\sqrt{c^2-4ab}}{a}\right)$. It was proved by Barrio et al. [12] that the equilibrium point $P_1$ of model (2) is always unstable and the equilibrium point $P_2$ is linearly stable if and only if the following criteria hold:

**(C1)** Either the inequalities $a \leq 1$, $c > 2a$ or the inequalities $1 < a < \sqrt{2}$, $2a < c < \frac{2a}{a^2-1}$ hold;

**(C2)** The inequality $b_1 \leq b < b_2$ is fulfilled, where

$$b_1 = \frac{a(2 - a^4 + ca^3 + 2a^2 - ca + c^2 + (c-a)\sqrt{a^6 - 4a^4 + 2ca^3 - 4a^2 + c^2})}{2(a^2 + 1)^2}$$

and

$$b_2 = \frac{c^2}{4a}.$$

Cascade of period-doubling bifurcations is one of the routes to chaos [13]-[15]. Period-doubling cascades can appear in several models used, for instance, in laser dynamics, chemical reactions, medicine, and neural networks [16]-[19]. In a parameter dependent continuous-time or discrete-time system, a period-doubling bifurcation occurs if a stable periodic solution loses its stability and is replaced by a new stable periodic solution of twice period when the parameter increases or decreases through a bifurcation value $\mu_j$, where $j \in \mathbb{N}$. In the case of a period-doubling cascade, this process takes place infinitely many times and as a result infinitely many unstable periodic solutions arise [15]. The sequence $\{\mu_j\}$ of period-doubling bifurcation values accumulates to a finite parameter value beyond which chaotic behavior emerges [13]-[15]. It was found by Feigenbaum [13, 20] that the value of the limit

$$\lim_{j \to \infty} \frac{\mu_{j+1} - \mu_j}{\mu_{j+2} - \mu_{j+1}} \tag{5}$$

is the universal constant $4.6692016091029 \dots$.

Another route to chaos observed in dynamical systems is intermittency, which was presented by Pomeau and Manneville [21, 22]. In this phenomenon, nearly periodic oscillations are interrupted by irregular bursts. In study [22], intermittency was classified as type I, type II, and type III in accordance with the Floquet multipliers. Types I, II, and III respectively correpond to the formation of cyclic-fold bifurcation, sub-critical Hopf bifurcation or Neimark-Sacker bifurcation, and sub-critical period-doubling bifurcation [23]. Some other kinds of intermittency are type X, type V, on-off, eyelet, crisis-induced, and ring [23]-[29]. The reader is referred to the studies [30]-[33] for some models possessing chaos through the intermittency route.

The main purpose of the present study is the demonstration of period-doubling route to chaos and intermittency in a perturbed Rössler system whose right-hand side is discontinuous under the assumptions (C1) and (C2). The model is obtained by perturbing system (2) with a piecewise constant function, and it is introduced in the next section.

The occurrence of period-doubling route to chaos in systems with delay and impulses were investigated in [34, 35], where continuous perturbations were utilized. Differently from the papers [34, 35], in this study we focus on the formation of period-doubling cascade in a Rössler model with discontinuous right-hand side. A way to generate on-off intermittency was demonstrated in [36] by using a two-dimensional subsystem driven by a Rössler system. In the present paper we also provide a way to generate continuous chaos through the intermittency route, but we make use of the logistic map to perturb a Rössler system. Additionally, type I intermittency is generated instead of on-off type. Furthermore, we would like to point out that the results of this study cannot be obtained from the ones of paper [37] because of two main reasons. Firstly, we utilize a model whose right-hand side is discontinuous, whereas coupled continuous-time systems are used in [37]. Secondly, the Lipschitz constant of the nonlinear term of the model investigated in this study is not sufficiently small as required in [37].

The rest of the present study is organized as follows. The next section is concerned with the explanation of the hybrid Rössler model whose dynamics will be investigated. The occurrence of period-doubling cascade as well as its extension among coupled Rössler systems is considered in Section 3. On the other hand, in Section 4, we show the presence of type I intermittency in the model. Moreover, the extension of intermittency is also demonstrated in that section. Finally, some concluding remarks are provided in Section 5.

**2. The Model**

Let $\tau = \{\tau_j\}$, $j = 0, 1, 2, 3, ...,$ be a sequence of real numbers such that

$$\tau_{j+1} = \tau_j + T \qquad (6)$$

for each $j$, where $T$ is a positive number. In this study, we investigate the dynamics of the non-autonomous perturbed Rössler system of the form

$$x' = -y - z + f_\mu(t, d, \tau)$$
$$y' = x + ay \qquad (7)$$
$$z' = b + z(x - c),$$

where $\mu$ is a real parameter and the positive constants $a, b$ and $c$ satisfy the assumptions (C1) and (C2) so that the equilibrium point $P_2$ of (2) described by (4) is linearly stable.

Here, the piecewise constant function $f_\mu(t, d, \tau)$ is defined by the equation

$$f_\mu(t, d, \tau) = d_j^\mu, \qquad (8)$$

for $t \in [\tau_j, \tau_{j+1})$, $j = 0, 1, 2, 3, ...,$ where the sequence $d = \{d_j^\mu\}$ is a solution of the logistic map

$$h_{j+1} = \mu h_j (1 - h_j), \quad j = 0, 1, 2, 3, \ldots, \tag{9}$$

and $d_0^\mu \in [0,1]$. For $0 < \mu \le 4$, the unit interval $[0,1]$ is invariant under the iterations of (9) [38]. Moreover, the sequence $d = \{d_j^\mu\}$ is $p$-periodic for some natural number $p$ if and only if the function $f_\mu(t, d, \tau)$ defined by (8) is periodic with period $pT$.

It is worth noting that the model under consideration is a hybrid one since its dynamics is governed by both the system of differential equations (7) and the discrete equation (9). For different values of the parameter $\mu$, the logistic map (9) possesses various dynamical phenomena such as homoclinic and heteroclinic orbits, period-doubling cascades, intermittency, and chaos in the senses of Li-Yorke, Devaney and Poincaré [39]-[42]. The logistic map was used in [43] to examplify continuous unpredictable motions in shunting inhibitory cellular neural networks. Moreover, the reader is referred to [44] for discontinuous unpredictable motions generated by impulsive systems.

In the dynamics of the map (9), period-doubling bifucations occur at the parameter values $\mu_1 = 3$, $\mu_2 = 3.4494896$, $\mu_3 = 3.5440903$, $\mu_4 = 3.5644073$, $\mu_5 = 3.5687594$, $\mu_6 = 3.5696916$, $\mu_7 = 3.5698913$, and so on [15]. The strictly increasing sequence $\{\mu_j\}, j \in \mathbb{N}$, has the limit

$$\mu_\infty = 3.569945, \tag{10}$$

beyond which chaos arises in (9) through the period-doubling route [15, 45].

For each fixed value of the parameter $\mu$ between $\mu_j$ and $\mu_{j+1}$, $j \in \mathbb{N}$, the logistic map (9) admits $2^j$ different $2^j$-periodic points [15]. For instance, for $\mu = 3.21$ the $2$-periodic points are $0.509304373857738$ and $0.802222105893041$, whereas for $\mu = 3.48$ the $4$-periodic points are $0.869426203367649$, $0.395064495323467$, $0.831680118706819$, and $0.487159280014113$.

A solution of the hybrid model (7) is described as follows.

**Definition 1.** ([46, 47]) Suppose that $\mu \in [0,4]$, $\tau_0 \in \mathbb{R}$, and $d_0^\mu \in [0,1]$ are fixed. A function $\phi: [\tau_0, \infty) \to \mathbb{R}^3$ is called a solution of system (7) if:

(i) $\phi(t)$ is continuous on $[\tau_0, \infty)$;

(ii) the derivative $\phi'(t)$ exists for every $t \ge \tau_0$ with the possible exception of the points $\tau_j, j = 0, 1, 2, 3, \ldots$, where one-sided derivatives exist;

(iii) $\phi(t)$ satisfies the differential equations in (7) for every $t \ge \tau_0$ except possibly at the points $\tau_j$, $j = 0, 1, 2, 3, \ldots$, and these differential equations hold for the left derivative of $\phi(t)$ at the points $\tau_j$.

The next section is devoted to the demonstration of a period-doubling cascade in model (7).

## 3. Period-Doubling Cascade and its Extension

Let us take $\tau_0 = 0$ and $T = 4$ in equation (6). We consider the hybrid Rössler model (7) with $a = 0.28$, $b = 1.43$, and $c = 1.57$. One can confirm that the criteria (C1) and (C2) are fulfilled with these values of $a$, $b$, and $c$ such that system (2) admits the equilibrium points $P_1$ and $P_2$ which are respectively defined by (3) and (4), where $q_1 = -4.462748904900945$ and

$q_2 = -1.144393952241912$. The equilibrium point $P_1$ is unstable whereas $P_2$ is linearly stable by Proposition 1 provided in paper [12].

For $\mu_j < \mu < \mu_{j+1}$, $j \in \mathbb{N}$, where $\{\mu_j\}$ is the sequence of period-doubling bifurcation values mentioned in the previous section, system (7) possesses $2^j$ different stable $2^j$ −periodic solutions whose trajectories coincide. Figure 1 depicts periodic and chaotic trajectories of system (7). The values 3.21, 3.48, and 3.55 of the parameter $\mu$ are respectively used in Figure 1 (a), (b), and (c). It is seen that as the parameter $\mu$ increases through the period-doubling bifurcation values $\mu_j$ of the map (9), a stable periodic solution of Rössler model (7) loses its stability and is replaced by another stable one of twice period. As a result, at the parameter values greater than $\mu_\infty$ chaos through period-doubling route can be observed. The trajectories shown in Figure 1 (a), (b), and (c) have the time periods 8, 16 and 32, respectively. To demonstrate the chaotic behavior, we show in Figure 1 (d) the irregular trajectory of (7) with $\mu = 3.59$ using the initial data $x(\tau_0) = -0.139$, $y(\tau_0) = -0.223$, $z(\tau_0) = 0.835$, and $d_0^\mu = 0.454$. The simulations shown in Figure 1 confirm the presence of period-doubling cascade in the dynamics of the Rössler model (7). On the other hand, Figure 2 (a), (b), (c) and (d) respectively depict the 2-dimensional projections of the trajectories shown in Figure 1 (a), (b), (c) and (d) on the $xy$-plane. Figure 2 also manifests the occurrence of a period-doubling cascade in (7).

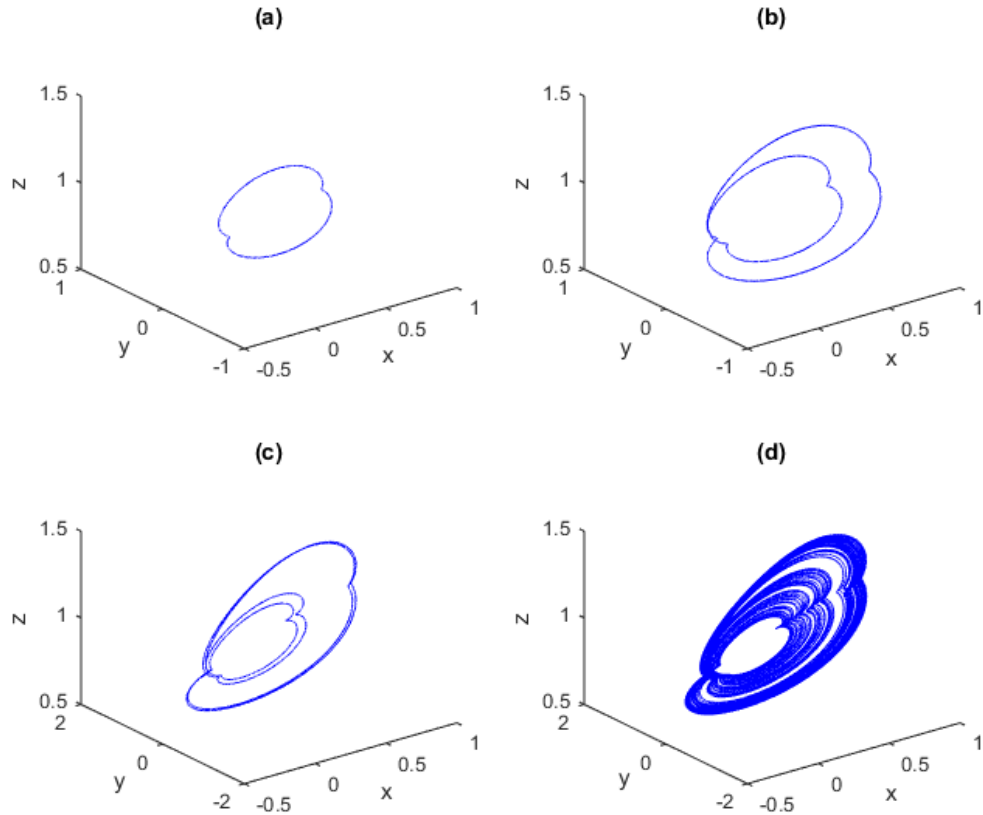

**Figure 1:** Period-doubling route to chaos in hybrid Rössler model (7). The values 3.21, 3.48, 3.55, and 3.59 for $\mu$ are respectively used in (a), (b), (c), and (d). The simulations in (a), (b), (c) depict periodic trajectories, and the simulation in (d) represents a chaotic trajectory.

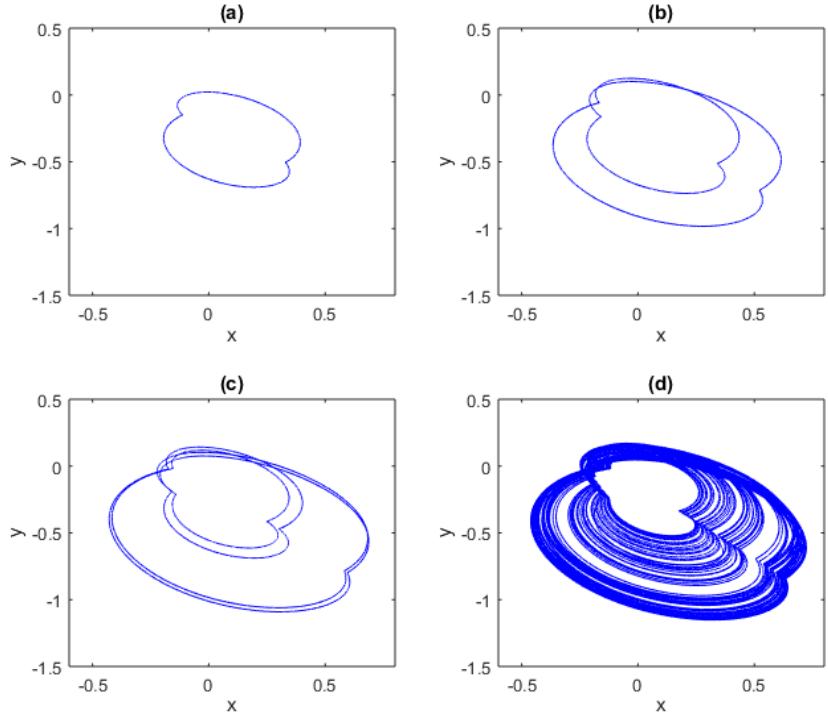

**Figure 2:** 2-dimensional projections of periodic and chaotic trajectories of hybrid Rössler model (7) on the $xy$-plane. The figure reveals the presence of a period-doubling cascade in the dynamics of (7).

It is remarkable that the period-doubling bifurcation values $\mu_j$, $j \in \mathbb{N}$, of system (7) are exactly the same with the logistic map (9). Therefore, the Feigenbaum universality is valid such that when the limit (5) is evaluated for system (7), the Feigenbaum's constant $4.6692016091029\ldots$ is achieved.

Now, in order to demonstrate the extension of period-doubling cascade among coupled Rössler systems, we set up the system

$$u' = -v - w + 0.005x(t)$$
$$v' = u + 0.62v + 0.009y(t) \tag{11}$$
$$w' = 1.29 + w(u - 1.86) + 0.007z(t),$$

where $(x(t), y(t), z(t))$ is a solution of (7) with $a = 0.28$, $b = 1.43$, $c = 1.57$, $\tau_0 = 0$, and $T = 4$.

We would like to emphasize that the counterparts of the assumptions (C1) and (C2) are fulfilled for the unperturbed system

$$u' = -v - w$$
$$v' = u + 0.62v \tag{12}$$
$$w' = 1.29 + w(u - 1.86),$$

such that it has the equilibrium points

$$Q_1 = (1.185147016443461,\ -1.911527445876551,\ 1.911527445876551)$$

and

$$Q_2 = (0.674852983556539, -1.088472554123449, 1.088472554123449),$$

where the former is unstable and the latter is linearly stable. Owing to the linear stability of the equilibrium point $Q_2$ of (12), it is possible to extend the period-doubling cascade which occurs in the dynamics of (7).

The projections of periodic and chaotic trajectories of the coupled system (7)-(11) on the $uvw$-space are represented in Figure 3. Likewise Figure 1, we again make use of the values 3.21, 3.48, 3.55, and 3.59 for $\mu$ in Figure 3 (a), (b), (c), and (d), respectively. Figure 3 reveals that the period-doubling cascade of (7) is extended by (11) such that a periodic solution $(x(t), y(t), z(t))$ of (7) used in the perturbation gives rise to the formation of a periodic solution of (11) with the same period, whereas if a solution $(x(t), y(t), z(t))$ of (7) chaotic, then (11) produces a solution with similar behaviour. The initial data utilized in Figure 3 (d) are $x(\tau_0) = -0.121$, $y(\tau_0) = -0.247$, $z(\tau_0) = 0.842$, $u(\tau_0) = 0.681$, $v(\tau_0) = -1.113$, $w(\tau_0) = 1.097$, and $d_0^\mu = 0.454$. Additionally, we represent in Figure 4 (a), (b), (c) and (d) the 2-dimensional projections of the trajectories depicted in Figure 3 (a), (b), (c) and (d) on the $uv$-plane, respectively. This figure also confirms the extension of period-doubling cascade of (7). Furthermore, one can conclude that the coupled system (7)-(11) admits the chaos through the period-doubling route.

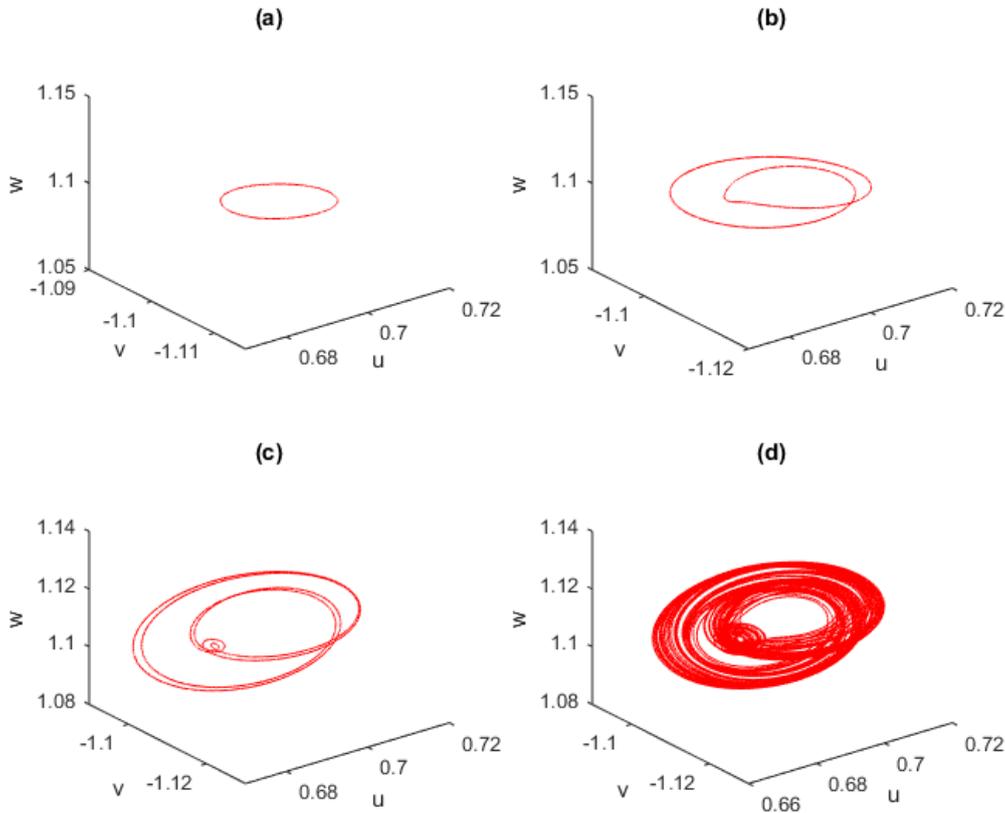

**Figure 3:** Extension of the period-doubling cascade of system (7) via coupling with (11). The figure shows the 3-dimensional projections of periodic and chaotic trajectories of the coupled system (7)-(11) on the $uvw$-space. The values 3.21, 3.48, 3.55, and 3.59 for $\mu$ are respectively used in (a), (b), (c), and (d).

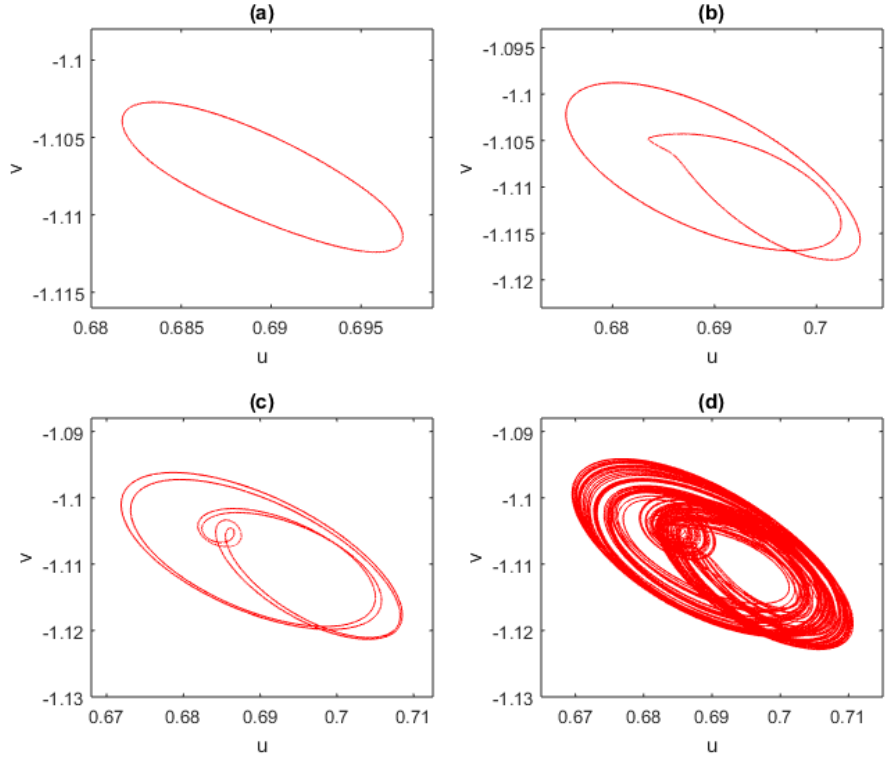

**Figure 4:** 2-dimensional projections of periodic and chaotic trajectories of the coupled system (7)-(11) on the $uv$-plane. The figure manifests period-doubling route to chaos in the coupled system (7)-(11).

### 4. Intermittency Route to Chaos

In this section, we take into account the hybrid Rössler model (7) with $\tau_0 = 0.4, T = 1.4, a = 0.47$, $b = 1.24$, and $c = 1.63$. With these choices of $a, b$, and $c$, the assumptions (C1) and (C2) are satisfied. The point $P_2$ given by (3) with $q_2 = -1.126913486239138$ is the linearly stable equilibrium of the unperturbed Rössler system (2), and equilibrium point $P_1$ given by (4) is unstable, where $q_1 = -2.341171620143841$.

For the values of the parameter $\mu$ slightly less than $1 + \sqrt{8}$ the logistic map admits chaos through the intermittency route which is of type I [30, 48]. Figure 5 shows the time series for all coordinates of (7) with $\mu = 3.82825$ corresponding to the initial data $x(\tau_0) = 0.836, y(\tau_0) = -0.467, z(\tau_0) = 1.797$, and $d_0^\mu = 0.4748$. The time interval $[150, 420]$ is used in the simulations. It is seen in Figure 5 that for $150 < t < 210$ and $360 < t < 420$ the behaviour is nearly periodic, whereas it is irregular for $210 < t < 360$. The interruption of the regular behavior by irregularity manifests the presence of chaos through the intermittency route.

Now, to demonstrate the extension of intermittency, we consider the coupled model (7)-(11), again with $\tau_0 = 0.4, T = 1.4, a = 0.47, b = 1.24$, and $c = 1.63$. We represent in Figure 6 the time series of the coordinates of (11) for $t \in [170, 430]$. It is displayed in the figure that the dynamics is nearly periodic for $170 < t < 210$ and $390 < t < 430$, while it is irregular for $210 < t < 390$. Therefore, one can conclude that system (11) also displays intermittency route to chaos, and accordingly, this type of chaotic behavior is extended.

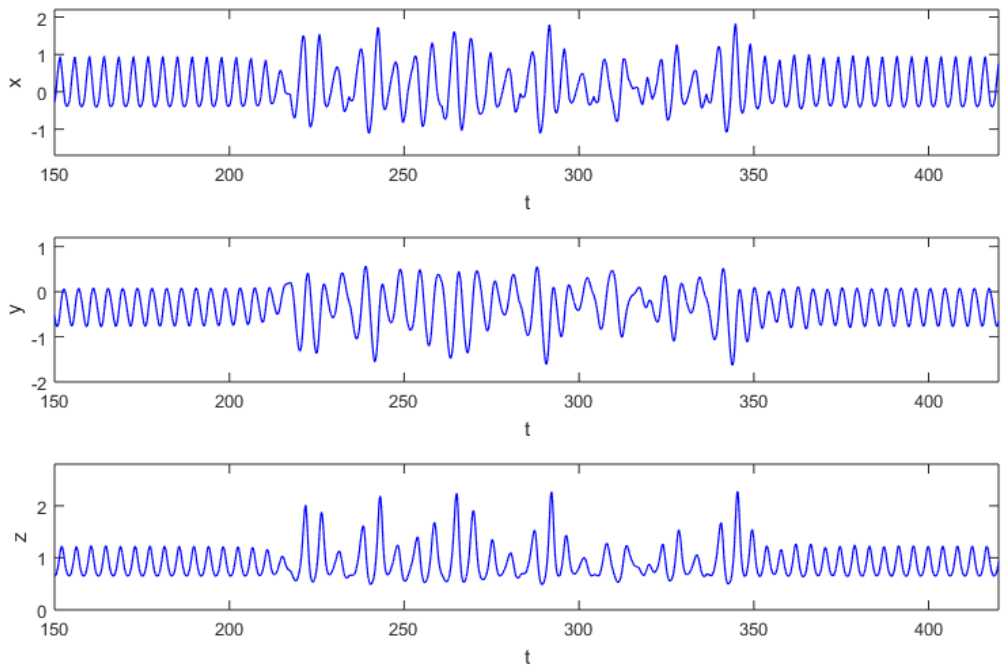

**Figure 5:** Intermittency route to chaos in the hybrid Rössler model (7). One can observe that nearly periodic behavior is interrupted by irregular bursts, and this approves the presence of intermittency.

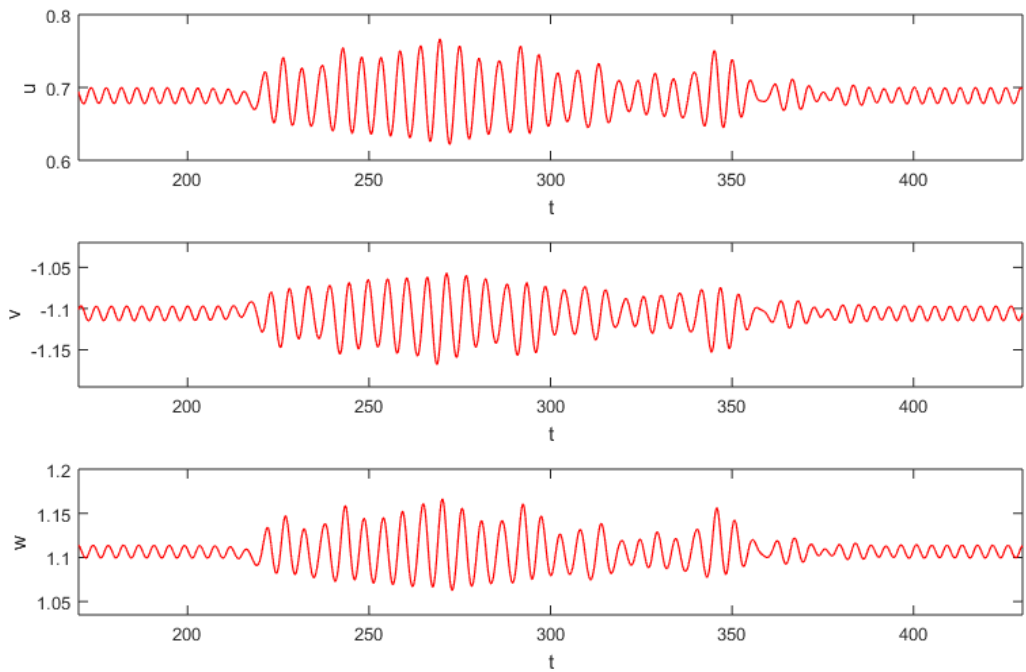

**Figure 6:** Extension of intermittency route to chaos. The figure shows the time-series of $u, v$, and $w$ coordinates of the coupled system (7)-(11). It is seen in the figure that irregular behavior takes place between regular ones. This reveals that the intermittency produced by system (7) is extended by (11).

## Conclusion

This study is devoted to the formation of period-doubling cascade and intermittency in a class of Rössler model whose right-hand side comprises a piecewise constant function. The model under investigation can be regarded as a hybrid one since its dynamics is governed by a system of differential equations (7) as well as the logistic map (9). We present a way to achieve continuous chaos from a discrete one. More precisely, the discrete chaos of the map (9) is transformed to continuous chaos via the flow of system (7). It is numerically demonstrated that likewise (9), system (7) possesses chaos through the period-doubling route for the values of the parameter $\mu$ beyond $\mu_\infty$. We showed that intermittency route to chaos also takes place for a certain parameter value. The main source of the chaotic behavior observed in (7) is the map (9) since in the absence of the perturbation the system has a linearly stable equilibrium point. The results manifest that discontinuous perturbations can give rise to the formation of chaos. The extension of chaotic behavior among coupled Rössler systems is demonstrated, too. Similar results can be also obtained if the parameter dependent perturbation term $f_\mu(t, d, \tau)$ defined by (8) is applied to the second or third differential equation in system (7).